\documentstyle[12pt,graphicx,pstricks,braket,amsmath,amssymb,enumitem,
                   float, epstopdf,pdfpages,cite,tabu,hyperref,subfigure, float,tikz]{article}
                    \usetikzlibrary{arrows,matrix,positioning}
                    
\numberwithin{equation}{section}
\DeclareMathOperator{\Tr}{Tr}
\DeclareMathOperator{\Res}{Res}
\DeclareMathOperator{\ad}{ad}
\begin{document}
	\def\AEF{A.E. Faraggi}
\def\JHEP#1#2#3{{JHEP} {\textbf #1}, (#2) #3}
\def\vol#1#2#3{{\bf {#1}} ({#2}) {#3}}
\def\NPB#1#2#3{{\it Nucl.\ Phys.}\/ {\bf B#1} (#2) #3}
\def\PLB#1#2#3{{\it Phys.\ Lett.}\/ {\bf B#1} (#2) #3}
\def\PRD#1#2#3{{\it Phys.\ Rev.}\/ {\bf D#1} (#2) #3}
\def\PRL#1#2#3{{\it Phys.\ Rev.\ Lett.}\/ {\bf #1} (#2) #3}
\def\PRT#1#2#3{{\it Phys.\ Rep.}\/ {\bf#1} (#2) #3}
\def\MODA#1#2#3{{\it Mod.\ Phys.\ Lett.}\/ {\bf A#1} (#2) #3}
\def\RMP#1#2#3{{\it Rev.\ Mod.\ Phys.}\/ {\bf #1} (#2) #3}
\def\IJMP#1#2#3{{\it Int.\ J.\ Mod.\ Phys.}\/ {\bf A#1} (#2) #3}
\def\nuvc#1#2#3{{\it Nuovo Cimento}\/ {\bf #1A} (#2) #3}
\def\RPP#1#2#3{{\it Rept.\ Prog.\ Phys.}\/ {\bf #1} (#2) #3}
\def\APJ#1#2#3{{\it Astrophys.\ J.}\/ {\bf #1} (#2) #3}
\def\APP#1#2#3{{\it Astropart.\ Phys.}\/ {\bf #1} (#2) #3}
\def\EJP#1#2#3{{\it Eur.\ Phys.\ Jour.}\/ {\bf C#1} (#2) #3}

\begin{titlepage}
\samepage{
\setcounter{page}{1}
\rightline{}
\vspace{1.5cm}

\begin{center}
 {\Large \bf Monodromic T-Branes And The $SO(10)_{GUT}$}
\end{center}

\begin{center}

{\large
Johar M. Ashfaque$^\spadesuit$\footnote{email address: jauhar@liv.ac.uk}
}\\
\vspace{1cm}
$^\spadesuit${\it  Dept.\ of Mathematical Sciences,
             University of Liverpool,
         Liverpool L69 7ZL, UK\\}
\end{center}

\begin{abstract}
T-branes, which are non-Abelian bound states of branes, were first introduced by Cecotti, Cordova, Heckman and Vafa \cite{Cecotti:2010bp}.  They are the
refined version of the monodromic branes that feature in the phenomenological F-theory models.  Here, we will be interested in the T-brane corresponding to the $Z_3$ monodromy which is used to break the $E_8$ gauge group to obtain the $SO(10)_{GUT}$. This extends the results of \cite{Cecotti:2010bp} to the case of $Z_3$ monodromic T-branes used to break the $E_8$ gauge group to $SO(10)\times SU(3)\times U(1)$ and compute the Yukawa coupling with the help of the residue formula. We conclude that the Yukawa coupling, ${\bf{10}}_{H}\cdot {\bf{16}}_{M}\cdot {\bf{16}}_{M}$, is non-zero for $E_7$, in complete agreement with \cite{Cecotti:2010bp}, but is zero for $E_8$. Furthermore, the case of $Z_2$ monodromic T-branes used to break the $E_8$ gauge group to $E_{6}\times SU(2)\times U(1)$, nothing interesting can be deduced by evaluating the Yukawa coupling ${\bf{27}}_{H}\cdot {\bf{27}}_{M}\cdot {\bf{27}}_{M}$  which is dependent on whether the MSSM fermion and electroweak Higgs fields can be included in the same ${\bf{27}}$ multiplet of a three-family $E_6$ GUT or assign the Higgs fields to a different ${\bf{27}}_{H}$ multiplet where only the Higgs doublets and singlets obtain the electroweak scale energy.
\end{abstract}
\smallskip}
\end{titlepage}
\tableofcontents 

\section{Introduction}
To break GUT symmetries within F-theory \cite{Vafa:1996xn, Morrison:1996na, Morrison:1996pp, Beasley:2008kw} in order to construct phenomenologically viable models, one can either make use of Wilson lines or introduce gauge fluxes with the key  ingredient being the seven-brane which wraps the four-dimensional internal subspace of the six internal directions of the compactification providing for each important element. The F-theory $E_{6}$ was discussed in \cite{Ashfaque:2016ydg} whilst the primary focus in \cite{Ashfaque:2016wfv}, was the $E_7$ gauge group obtained in this setting.\footnote{Refer to Appendix \ref{FTH} for details.}

Here, however, the focus is on T-branes, or ``triangular branes," which are novel non-Abelian bound states of branes characterized by the condition that on some loci the Higgs field is upper triangular and indeed it can be seen to be the case in \cite{Cecotti:2010bp} where $\langle \Phi \rangle$ is upper triangular on some locus. This approach deals with the spectral equation $$P_{\Phi}(z)=\det (z-\Phi) =0$$
which when $\Phi$ belongs to the CSA is equivalent to stating that
$$\prod_{i} (z-\lambda_{i})=0$$
where $\lambda_{i}$ are the eigenvalues of $\Phi$ and they denote the directions of the intersecting branes. In the case of non-diagonalizable Higgs fields, the monodromy group is now encoded in the form of the spectral equation. Such configurations are of particular interest when systems of $7$-branes are considered \cite{res}. When considering such a background profile for $\Phi$, it has to be made sure that the following equations of motion are satisfied which read 
\begin{eqnarray*}
	\overline{\partial}_{A}\Phi &=& 0\\
	F^{(0,2)}_{A}	5&=&0
\end{eqnarray*}
for the $F$-term equations and for the $D$-term equation we have that 
$$\omega\wedge F_{A} +\frac{i}{2}[\Phi^{\dag},\Phi]=0.$$

In this paper, the aim is to compute the Yukawa coupling with the help of the residue formula where the required interaction terms are of the form
$${\bf{10}}_{H}\cdot {\bf{16}}_{M}\cdot {\bf{16}}_{M}.$$ 
The present work extends the results of \cite{Cecotti:2010bp} to the case of
$Z_3$ monodromic T-branes used to break the $E_8$ gauge group to $SO(10)\times SU(3)\times U(1)$ where the case of $Z_2$ monodromic T-branes used to break the $E_7$ gauge group to $SO(10)\times SU(2)\times U(1)$ is reviewed beforehand. We will show that the Yukawa coupling, ${\bf{10}}_{H}\cdot {\bf{16}}_{M}\cdot {\bf{16}}_{M}$, is non-zero for $E_7$ but for $E_8$ is zero. Moreover, the case of $Z_2$ monodromic T-branes used to break the $E_8$ gauge group to $E_{6}\times SU(2)\times U(1)$, nothing interesting can be deduced by evaluating the Yukawa coupling ${\bf{27}}_{H}\cdot {\bf{27}}_{M}\cdot {\bf{27}}_{M}$ which depends on whether the MSSM fermion and electroweak Higgs fields can be included in the same ${\bf{27}}$ multiplet of a three-family $E_6$ GUT or assign the Higgs fields to a different ${\bf{27}}_{H}$ multiplet where only the Higgs doublets and singlets obtain the electroweak scale energy. This work can also seen as an extension to \cite{Ashfaque:2016wfv} in pursuit of obtaining $SO(10)$ GUT symmetry especially the Flipped $SO(10)$ from various string theoretic constructions.  

\section{The $Z_{2}$ Monodromy} \label{E7}
\subsection{Review of $SU(2)$ Field}
Let us consider the spectral equation for an $SU (2)$ field along the lines of \cite{Cecotti:2010bp}:
$$P_{\Phi}(z)=z^{2}-x$$
for which there is a $Z_{2}$ monodromy.
In the holomorphic gauge the Higgs field is
\[ \left( \begin{array}{cc}
0 & 1 \\
x & 0 \end{array} \right)\]
which is an intermediate case between a diagonal background and a nilpotent Higgs field  
\[ \left( \begin{array}{cc}
0 & 1  \\
0 & 0 \end{array} \right).\]
We promote this to the unitary gauge by use of a positive diagonal matrix having unit determinant 
\[ \left( \begin{array}{cc}
e^{f_{1}}& 0 \\
0 & e^{f_{2}} \end{array} \right)\]
with the constraint 
$$\sum_{i}f_{i}=0$$
where $f_{i}$ are real. The $D$-term equation
$$\omega\wedge F_{A} +\frac{i}{2}[\Phi^{\dag},\Phi]=0$$
is now replaced by the $SU(2)$ Toda equation in two complex variables	
$$\Delta f_{i}=C_{ij} e^{f_{j}}$$
where $C_{ij}$ is the Cartan matrix of $SU(2)$.

\subsection{The Brane Recombination}
Infinitesimal perturbations to the  holomorphic Higgs field are considered  of the form
$$\varphi =\ad_{\Phi}(\xi)+h$$ and then seen which can be gauged away to zero by the $U(2)$ transformations by way of deforming the theory via $SU(2)$ Higgs VEV 
\[ \Phi=\left( \begin{array}{cc}
0 & 1  \\
x & 0 \end{array} \right).\]
After gauge fixing the most general perturbation that can be made is given by
\[ \varphi=\left( \begin{array}{cc}
\frac{1}{2}\alpha(x,y)& 0 \\\beta(x,y)& \frac{1}{2}\alpha(x,y)\end{array} \right).\]
With such a perturbation at hand, the spectral equation can seen to be deformed by the $SU(2)$ Higgs VEV as
$$P_{\Phi}(z)=z^{2}-x\rightarrow \bigg(\bigg(z-\frac{1}{2}\alpha(x,y)\bigg)^{2}-(x+\beta(x,y)\bigg).$$
Now changing coordinates yields
$$\tilde{z}-(\tilde{x}\alpha(\tilde{x}^{2},\tilde{y})+\beta(\tilde{x}^{2},\tilde{y})).$$
This is interpreted as the three $D7$-branes recombining into one.

Now starting with the flat K\"ahler metric 
$$\omega =\frac{i}{2}(dx\wedge d\overline{x}+dy\wedge d\overline{y}+dz\wedge d\overline{z})$$
changing to the new coordinates and noting that $x=\tilde{x}^{2}$, we have 
$$\omega = \frac{i}{2}\big(\big(1+4|\tilde{x}|^{2}\big)d\tilde{x}\wedge d\overline{\tilde{x}}+d\tilde{y}\wedge d\overline{\tilde{y}}\big)$$
and therefore the recombined $D7$-brane is indeed curved.
\subsection{$E_{7}\rightarrow SO(10)\times SU(2)\times U(1)$}
Here $SU(2)\times U(1)$ Higgs field is used that preserves an unbroken $SO(10)$:
\[ \Phi=\left( \begin{array}{cc}
0 & 1  \\
x & 0 \end{array} \right)\oplus (y).\]
The adjoint of $E_7$ decomposes under this breaking as
\begin{eqnarray*}
	\bf{133}&\rightarrow& (\bf{1},\bf{1})_{0}\oplus (\bf{1},\bf{3})_{0}\oplus (\bf{45},\bf{1})_{0}\oplus (\bf{10}, \bf{1})_{2}\oplus (\bf{10},\bf{{1}})_{-2}\oplus(\bf{16},\bf{2})_{-1}\oplus(\bf{\overline{16}},\bf{2})_{1} 
\end{eqnarray*}\\
where the $U(1)$ generator is in parenthesis. We note that the required interaction terms are of the form
$${\bf{10}}_{H}\cdot {\bf{16}}_{M}\cdot {\bf{16}}_{M}.$$ 
 We can readily identify $(\bf{10}, \bf{1})_{2}$ as the ${\bf{10}}_{H}$ and $(\bf{16},\bf{2})_{-1}$ as ${\bf{16}}_{M}$ with
\[ \varphi_{{\bf{16}}_{M}}=\left( \begin{array}{c}
\varphi_{{\bf{16}}+}\\ \varphi_{{\bf{16}}-} \end{array} \right)\]
where the corresponding matter curve is
$$f=y^{2}-x.$$
Note that the one component of the spinor doublet, namely $\varphi_{{{\bf{16}}}+}$ is gauge equivalent to zero.

The trace in the adjoint of ${\mathfrak{e}}_7$ produces the following invariant tensors
$$\Tr ([t_{{\bf{10}},i}, t^{M}_{{\bf{16}},\alpha}]t^{N}_{{\bf{16}},\beta}) \propto (C\Gamma_{i})_{\alpha\beta}\epsilon^{MN}$$
which is generated group theoretically by contraction with a $\Gamma$ matrix of the $SO(10)$ Clifford algebra, where $\alpha$, $\beta$ are spinor indices,
$i$ is a vector index, and $C$ denotes the standard charge conjugation matrix. 

The Yukawa coupling can now be evaluated simply as 
\begin{eqnarray*}
	W_{{\bf{10}}\cdot {\bf{16}}\cdot {\bf{16}}} &=& \Res_{(0,0)}\bigg[\frac{\Tr([\eta_{\bf{10}},\eta_{\bf{16}}]\varphi_{\bf{16}})}{(y)(y^{2}-x)}\bigg]\\
	&=& \Res_{(0,0)}\bigg[\frac{(C\Gamma_{i})_{\alpha\beta}\varphi^{\alpha}_{\bf{16}-} \varphi^{\beta}_{\bf{16}-} \varphi^{i}_{\bf{10}}}{(x)(y)}\bigg]
\end{eqnarray*}
where $$\eta_{{\bf{10}}_H} = \frac{1}{2}\varphi_{{\bf{10}}_H}$$ and 
\
\[ \eta_{\bf{16}}=-\left( \begin{array}{c}
\varphi_{{\bf{16}}-}\\ y\varphi_{{\bf{16}}-} \end{array} \right).\]
This coupling requires a single field, $\varphi_{{\bf{16}}-}$, to participate twice in the trilinear Yukawa coupling and is known to give mass to exactly one generation of SM matter ${\bf{16}}$'s.

\section{The $Z_{3}$ Monodromy}
\subsection{Review of $SU(3)$}
We follow \cite{Cecotti:2010bp,Chiou:2011js} to outline some of the key ideas to paint the picture. Let us begin by considering the spectral equation for an $SU (3)$ field:
$$P_{\Phi}(z)=z^{3}-x$$
for which there is a $Z_{3}$ monodromy.

In the holomorphic gauge the Higgs field is
\[ \left( \begin{array}{ccc}
0 & 1 & 0 \\
0 & 0 & 1 \\
x & 0 & 0\end{array} \right)\]
which is an intermediate case between a diagonal background and a nilpotent Higgs field  
\[ \left( \begin{array}{ccc}
0 & 1 & 0 \\
0 & 0 & 1 \\
0 & 0 & 0\end{array} \right).\]
We promote this to the unitary gauge by use of a positive diagonal matrix having unit determinant 
\[ \left( \begin{array}{ccc}
e^{f_{1}}& 0 & 0 \\
0 & e^{f_{2}} & 0 \\
0 & 0 & e^{f_{3}}\end{array} \right)\]
with the constraint 
$$\sum_{i}f_{i}=0$$
where $f_{i}$ are real. The $D$-term equation
$$\omega\wedge F_{A} +\frac{i}{2}[\Phi^{\dag},\Phi]=0$$
is now replaced by the $SU(3)$ Toda equation in two complex variables	
$$\Delta f_{i}=C_{ij} e^{f_{j}}$$
where $C_{ij}$ is the Cartan matrix of $SU(3)$ which is given by
\[ \left( \begin{array}{cc}
\,\,\,\,2& -1 \\ -1 & \,\,\,\,2
\end{array} \right).\]
The components for the unitary transformation for the nilpotent Higgs field $\Phi$ satisfy
\begin{eqnarray*}
\partial \overline{\partial}f_{1}&=&2e^{f_{1}} -e^{f_{2}},\\
\partial \overline{\partial}f_{2} &=& -e^{f_{1}}+2e^{f_{2}}.
\end{eqnarray*}
\subsection{The Brane Recombination}
Infinitesimal perturbations to the  holomorphic Higgs field are considered  of the form
$$\varphi =\ad_{\Phi}(\xi)+h$$ and then seen which can be gauged away to zero by the $U(3)$ transformations by way of deforming the theory via $SU(3)$ Higgs VEV 
\[ \Phi=\left( \begin{array}{ccc}
0 & 1 & 0 \\
0 & 0 & 1 \\
x & 0 & 0\end{array} \right).\]
After gauge fixing, as was shown in \cite{Chiou:2011js}, the most general perturbation that can be made is given by
\[ \varphi=\left( \begin{array}{ccc}
 \frac{1}{3}\alpha(x,y)& 0 & 0 \\
0 & \frac{1}{3}\alpha(x,y) & 0 \\
\gamma(x,y) & \beta(x,y) & \frac{1}{3}\alpha(x,y)\end{array} \right).\]
With such a perturbation at hand, the spectral equation can seen to be deformed by the $SU(3)$ Higgs VEV as
$$P_{\Phi}(z)=z^{3}-x\rightarrow \bigg(z-\frac{1}{3}\alpha(x,y)\bigg)\bigg(\bigg(z-\frac{1}{3}\alpha(x,y)\bigg)^{2}-\beta(x,y)\bigg)-(x+\gamma(x,y))$$
which to first order reads
$$z^{3}-z^{2}\alpha(x,y)-z\beta(x,y)-x-\gamma(x,y).$$
Now changing coordinates to 
$$(\tilde{x},\tilde{y},\tilde{z})=(z,y,P_{\Phi}(z))$$
yields
$$\tilde{z}-(\tilde{x}^{2}\alpha(\tilde{x}^{3},\tilde{y})+\tilde{x}\beta(\tilde{x}^{3},\tilde{y})+\gamma(\tilde{x}^{3},\tilde{y}) ).$$
This is interpreted as the three $D7$-branes recombining into one. 

Now starting with the flat K\"ahler metric 
$$\omega =\frac{i}{2}(dx\wedge d\overline{x}+dy\wedge d\overline{y}+dz\wedge d\overline{z})$$
changing to the new coordinates and noting that $x=\tilde{x}^{3}$, we have 
$$\omega = \frac{i}{2}\big(\big(1+9|\tilde{x}|^{4}\big)d\tilde{x}\wedge d\overline{\tilde{x}}+d\tilde{y}\wedge d\overline{\tilde{y}}\big)$$
and therefore the recombined $D7$-brane is indeed curved.
\subsection{$E_{8}\rightarrow SO(10)\times SU(3)\times U(1)$}
Here $SU(3)\times U(1)$ Higgs field is used that preserves an unbroken $SO(10)$:
\[ \Phi=\left( \begin{array}{ccc}
0 & 1 & 0 \\
0 & 0 & 1 \\
x & 0 & 0\end{array} \right)\oplus (y).\]
The adjoint of $E_8$, following \cite{Slansky:1981yr}, decomposes as
\begin{eqnarray*}
\bf{248}&\rightarrow& (\bf{1},\bf{1})_{0}\oplus (\bf{1},\bf{8})_{0}\oplus (\bf{45},\bf{1})_{0}\oplus (\bf{1},\bf{3})_{-4}\oplus (\bf{1},\bf{\overline{3}})_{4}\oplus (\bf{10}, \bf{3})_{2}\oplus (\bf{10},\bf{\overline{3}})_{-2}\oplus\\
&& (\bf{16},\bf{1})_{3}\oplus(\bf{\overline{16}},\bf{1})_{-3} \oplus  (\bf{16},\bf{3})_{-1}\oplus(\bf{\overline{16}},\bf{\overline{3}})_{1}
\end{eqnarray*}\\
where the $U(1)$ generator is in parenthesis. We note that the required interaction terms are of the form
$${\bf{10}}_{H}\cdot {\bf{16}}_{M}\cdot {\bf{16}}_{M}.$$ 
We can readily identify $(\bf{10}, \bf{3})_{2}$ as the ${\bf{10}}_{H}$ and $(\bf{16},\bf{3})_{-1}$ as ${\bf{16}}_{M}$. The ${\bf{16}}_{M}$ is in the fundamental of the $SU(3)$
\[ \varphi_{{\bf{16}}_{M}}=\left( \begin{array}{c}
\varphi_{{\bf{16}}_{M}}^{1} \\
\varphi_{{\bf{16}}_{M}}^{2}  \\
\varphi_{{\bf{16}}_{M}}^{3}\end{array} \right).\]
 
The torsion equation can be solved using the adjugate matrix 
\[ \eta_{{\bf{16}}_{M}}=\underbrace{\left( \begin{array}{ccc}
4y^{2}&x&-2xy\\
-2y&4y^{2}&x\\
1&-2y&4y^{2}\end{array} \right)}_{A}\left( \begin{array}{c}
\varphi_{{\bf{16}}_M}^{1}\\
0\\0
\end{array} \right)  =  \left( \begin{array}{c}
4y^{2}\varphi_{{\bf{16}}_M}^{1}\\
-2y\varphi_{{\bf{16}}_M}^{1}\\\varphi_{{\bf{16}}_M}^{1}
\end{array} \right) .\]
The ${\bf{10}}_{H}$ transforms in the fundamental of $SU(3)$ as well and as a result replacing $y$ with -$2y$ the solution of the torsion equation is found to be 
\[ \eta_{{\bf{10}}_{H}}=\underbrace{\left( \begin{array}{ccc}
16y^{2}&x&4xy\\
4y&16y^{2}&x\\
1&4y&16y^{2}\end{array} \right)}_{B}\left( \begin{array}{c}
\varphi_{{\bf{10}}_H}^{1}\\
0\\0
\end{array} \right)  =  \left( \begin{array}{c}
16y^{2}\varphi_{{\bf{10}}_H}^{1}\\
4y\varphi_{{\bf{10}}_H}^{1}\\\varphi_{{\bf{10}}_H}^{1}
\end{array} \right). \]

Using that the matter curve corresponding to $A$ and $B$ are
$$f_A = 8y^{3}+x,\qquad\qquad f_B= 64y^{3}-x$$

and again using the fact that the trace in the adjoint of ${\mathfrak{e}}_{8}$ is  
$$\Tr ([t_{{\bf{10}},i}^{A}, t_{{\bf{16}},\alpha}^{B}]t_{{\bf{16}},\beta}^{C}) \propto (C\Gamma_{i})_{\alpha\beta}\epsilon^{ABC}$$
allowing the Yukawa coupling to be evaluated simply as 
\begin{eqnarray*}
	W_{{\bf{10}}\cdot {\bf{16}}\cdot {\bf{16}}} &=& 0
\end{eqnarray*}
as was expected since it was noted in \cite{Chiou:2011js} that for $n\geq 2$ and arbitrary $a$ and $b$, matter curves of the form 
$$f=ay^{n}+bx$$
will always yield zero for the computation of the trilinear Yukawa coupling. 

\section{Discussion And Conclusion}
The low-energy string-derived model of \cite{frzprime} was 
constructed in the free fermionic formulation \cite{fff} of the four-dimensional heterotic string in which the space-time vector bosons are obtained solely from the untwisted sector and generate the observable and hidden gauge symmetries:
\begin{eqnarray*}
	{\rm observable} ~&: &SO(6)\times SO(4) \times 
	\sum_{i=1}^{3}U(1)_i \nonumber\\
	{\rm hidden}     ~&: &SO(4)^2\times SO(8)~.~~~~~~~~~~~~~~~~~~~~~~~\nonumber
\end{eqnarray*}
where the $E_6$ combination being
\begin{eqnarray*}
	U(1)_\zeta = \sum_{i=1}^{3}U(1)_i ~,
	\label{u1zeta}
\end{eqnarray*}
which is anomaly free whereas the orthogonal combinations of $U(1)_{1,2,3}$
are anomalous.

Motivated by such string-derived low-energy effective models the Flipped $SO(10)$ was derived from the F-theory $E_{6}$ as was discussed in \cite{Ashfaque:2016ydg}  and investigated further in \cite{Ashfaque:2016wfv}, where the gauge group was $E_{7}$. Another possibility, that was explored in \cite{Ashfaque:2016wfv}, was that of nonperturbative heterotic vacua arising from the Hor\v{a}va-Witten theory.\footnote{See Appendix \ref{HWT} for the rules which allow the construction of realistic, viable vacua with $E_{6}$ GUT symmetry where the base manifold is taken to be the Hirzebruch surfaces.} The space of solutions of type A contains exactly one vacua over the Hirzebruch surfaces for any allowed value of $r$ for $6\leq s\leq 24$ and the corresponding appropriate choice for $\lambda$ with  
$$s\,\,\text{even},\,\,e-r\,\,\text{even},\,\,\lambda=\pm1, \pm3. $$
\begin{table}[H]
	\begin{center}
		\begin{tabular}{|c|c|}
			\hline
			$s$&$e(r; \lambda)$\\
			\hline
			&\\
			6&$3r+6+\frac{1}{\lambda}\,\,\,\,\in\mathbb{Z}$\\
			&\\
			&$\lambda = \pm 1$\\
			\hline
		\end{tabular}
			\end{center}
\end{table}
\noindent whereas the space of solutions of Type B 
$$r\,\,\text{even},\,\,\lambda = \pm\frac{1}{2},\pm \frac{3}{2}.$$
are given by 

$$s=6,\quad e\bigg(r;\lambda=\pm\frac{1}{2}\bigg)=3r+6+\frac{1}{\lambda}=3r+6\pm2.$$

In this paper, the aim was to compute the Yukawa coupling with the help of the residue formula where the required interaction terms are of the form
$${\bf{10}}_{H}\cdot {\bf{16}}_{M}\cdot {\bf{16}}_{M}$$ 
serving as an extension to the results of \cite{Cecotti:2010bp} to the case of
$Z_3$ monodromic T-branes used to break the $E_8$ gauge group to $SO(10)\times SU(3)\times U(1)$.  We conclude that the Yukawa coupling, ${\bf{10}}_{H}\cdot {\bf{16}}_{M}\cdot {\bf{16}}_{M}$, is non-zero for $E_7$, in complete agreement with \cite{Cecotti:2010bp}, but for $E_8$ is zero. Furthermore, the case of $Z_2$ monodromic T-branes used to break the $E_8$ gauge group to $E_{6}\times SU(2)\times U(1)$, nothing interesting can be deduced by evaluating the Yukawa coupling ${\bf{27}}_{H}\cdot {\bf{27}}_{M}\cdot {\bf{27}}_{M}$ which depends on whether the MSSM fermion and electroweak Higgs fields can be included in the same ${\bf{27}}$ multiplet of a three-family $E_6$ or assign the Higgs fields to a different ${\bf{27}}_{H}$ multiplet where only the Higgs doublets and singlets obtain the electroweak scale energy.  The work presented here can also be viewed as an extension to \cite{Ashfaque:2016wfv} in pursuit of obtaining $SO(10)$ GUT symmetry especially the Flipped $SO(10)$ from various string theoretic constructions.  

\appendix 
\section{The F-Theory Construct} \label{FTH}

\subsection{$E_6$}
$$E_{8}\supset E_{6}\times SU(3)_{\perp}$$
with 
$${\bf{248}} \rightarrow ({\bf{78}},{\bf{1}})+({\bf{1}},{\bf{8}})+({\bf{27}},{\bf{3}})+(\overline{{\bf{27}}},\overline{{\bf{3}}})$$
where the inhomogeneous Tate form for $E_6$ is given by
$$x^{3}-y^{2}+b_{1}xyz+b_{2}x^{2}z^{2}+b_{3}yz^{2}+b_{4}xz^{3}+b_{6}z^{5}=0.$$
In the spectral cover approach the $E_{6}$ representations are distinguished by the weights $t_{1,2,3}$ of the $SU(3)_{\perp}$ Cartan subalgebra subject to the traceless condition 
$$\sum_{i=1}^{3}t_{i}=0$$
while the $SU(3)_{\perp}$ adjoint decomposes into singlets. 
\subsection{$E_{7}$}
$$E_{8}\supset E_{7}\times SU(2)_{\perp}$$
with 
$${\bf{248}} \rightarrow ({\bf{133}},{\bf{1}})\oplus({\bf{1}},{\bf{3}})\oplus({\bf{56}},{\bf{2}})$$
where the inhomogeneous Tate form for $E_7$ is given by
$$x^{3}-y^{2}+b_{1}xyz+b_{2}x^{2}z^{2}+b_{3}yz^{3}+b_{4}xz^{3}+b_{6}z^{5}=0.$$
\subsection{The Gauge Enhancements}

\subsubsection{ $SO(10)$ }
\begin{eqnarray*}\Delta &=& -16b_{2}^{3}b_{3}^{2}z^{7}+\big(-27b_{3}^{4}-8b_{1}^{2}b_{2}^{2}b_{3}^{2}+72b_{2}b_{4}b_{3}^{2}\\&&\qquad\qquad\quad+4b_{1}b_{2}(9b_{3}^{2}+4b_{2}b_{4})b_{3}+16b_{2}^{2}(b_{4}^{2}-4b_{2}b_{6})\big)z^{8} \\&&\qquad \qquad\qquad \qquad\qquad \qquad\qquad \qquad \qquad\qquad\qquad\quad + \mathcal{O}(\mathit{z}^9)\\
	&=& z^{7} \big[-16b_{2}^{3}b_{3}^{2}+\big(-27b_{3}^{4}-8b_{1}^{2}b_{2}^{2}b_{3}^{2}+72b_{2}b_{4}b_{3}^{2}\\&&\qquad\qquad\quad+4b_{1}b_{2}(9b_{3}^{2}+4b_{2}b_{4})b_{3}+16b_{2}^{2}(b_{4}^{2}-4b_{2}b_{6})\big)z \\&&\qquad \qquad\qquad \qquad\qquad \qquad\qquad \qquad \qquad\qquad\qquad\quad + \mathcal{O}(\mathit{z}^2) \big]
\end{eqnarray*}

\begin{center}
	\begin{tabular}{|l|l|l|l|l|}
		\hline
		&$\deg(\Delta)$&Type&Gauge Group&Object Equation\\
		\hline
		GUT&$7$&$D_{5}$&$SO(10)$&$S:z=0$\\
		\hline
		Matter &$8$&$D_{6}$&$SO(12)$&$P_{10}: b_{3}=0 $\\
		Curve&&&&\\
		\hline
		Matter&$8$&$E_{6}$&$E_{6}$&$P_{16}: b_{2}=0 $\\
		Curve&&&&\\
		\hline
		Yukawa &$9$&$E_{7}$&$E_{7}$&$b_{2}=b_{3}=0$ \\
		Points&&&&$ b_{3}=b_{4}^{2}-4b_{2}b_{6}=0$\\
		\hline
	\end{tabular}
\end{center}

\subsubsection{$E_7$ }

\begin{eqnarray*}
	\Delta&=& z^{9} \big[-1024b_{4}^{3}+\big(((b_{1}^{2}+4b_{2})^{2}-96b_{1}b_{3})b_{4}^{2}\\&&\qquad\qquad\quad+72(b_{1}^{2}+4b_{2})b_{4}b_{6}-432b_{6}^{2}\big)z  + {\mathcal{O}}({\mathit{z}}^2) \big]
\end{eqnarray*}

\begin{table}[H]
	\begin{center}\begin{tabular}{|l|l|l|l|l|}
			\hline
			&$\deg(\Delta)$&Type&Gauge Group&Object Equation\\
			\hline
			GUT&$9$&$E_{7}$&$E_7$&$S:z=0$\\
			\hline
			Matter&$10$&$E_{8}$&$E_{8}$&$b_{4}=0$\\
			Curve&&&&\\
			\hline
		\end{tabular}
	\end{center}
\end{table}

\section{The Hirzebruch Surfaces $F_{r}$} \label{HWT}
The rules for constructing realistic, viable vacua with $E_{6}$ GUT symmetry where the base manifold $B$ are taken to be the Hirzebruch surfaces, $F_{r}$ are given. We arrive at the following conditions modified for the $E_{6}$ observable gauge group: 

\subsection{The Semistability Condition}
The semistability condition offers a choice: either 
$$\lambda\in \mathbb{Z}$$
and  $$s\,\,\text{even},\,\,e-r\,\,\text{even}$$
or 
$$\lambda = \frac{2m-1}{2},\,\,m \,\,\text{even},\quad r\,\,\text{even}.$$ 
\subsection{The Involution Conditions}
The involution conditions are 
$$\sum_{i}\kappa_{i}=\eta \cdot c_{1}(B=F_{r}) = 2e+2s-rs.$$
\subsection{The Effectiveness Condition}
The effectiveness condition boils down to
$$s \leq 24, \,\, \text{and}\,\,12r+24 \geq e$$
with 
$$\sum_{i}\kappa_{i}^{2}\leq 100+\frac{9}{4\lambda}-9\lambda$$
and 
$$\sum_{i}\kappa_{i}^{2}\leq 4+\frac{9}{4\lambda}-9\lambda+\sum_{i}\kappa_{i}.$$
\subsection{The Commutant Condition}
The commutant condition for $E_{6}$ becomes
$$\eta\geq 3c_{1}$$
which implies that
$$s\geq 6,\,\,\text{and}\,\,e\geq 3r+6.$$
\subsection{The Three Family Condition}
The three family condition reads 
$$-rs^{2}+3rs +2es -6e-6s =\frac{6}{\lambda}.$$
Solving the three family condition for $e$ assuming that the value of $s$ is known leads to
$$e(r;\lambda)  = \frac{1}{2s-6}\bigg(rs^{2}-3rs+6s+\frac{6}{\lambda}\bigg).$$

\end{document}